\documentclass[a4paper, conference]{IEEEtran}

\usepackage[utf8]{inputenc}
\usepackage{amssymb}
\usepackage{amsmath,bm}
\usepackage{mathrsfs}
\usepackage{xcolor}
\usepackage{scalerel}
\usepackage{theorem}
\usepackage{cite}
\usepackage{url}
\usepackage{array,multirow}
\usepackage{pgfplots}
\usetikzlibrary{plotmarks,patterns}
\usepackage[ruled,vlined,titlenumbered,linesnumbered]{algorithm2e}
\usepackage[nolist]{acronym}
\usepackage[innermargin=0.5in,outermargin=0.5in,top=0.6in,bottom=1in]{geometry}

\newtheorem{theorem}{Theorem}
\newtheorem{definition}{Definition}
\newtheorem{lemma}{Lemma}

\newtheorem{problem}{Problem}
\newtheorem{remark}{Remark}
\newtheorem{proposition}{Proposition}

\newcommand{\Fqm}{\ensuremath{\mathbb F_{q^m}}}

\newcommand{\Fqs}{\ensuremath{\mathbb F_{q^s}}}
\newcommand{\Fq}{\ensuremath{\mathbb F_{q}}}

\newcommand{\F}{\ensuremath{\mathbb F}}
\newcommand{\ZZ}{\ensuremath{\mathbb{Z}}}
\newcommand{\set}[1]{\ensuremath{\mathcal{#1}}}

\newcommand{\intervallincl}[2]{\ensuremath{[#1,#2]}}

\newcommand{\aut}{\ensuremath{\sigma}}

\newcommand{\SkewPolyringZeroDer}{\ensuremath{\Fqm[x,\aut]}}

\newcommand{\opev}[3]{\ensuremath{{#1}(#2)_{#3}}}

\newcommand{\op}[2]{\ensuremath{\mathcal{D}_{#1}(#2)}}
\newcommand{\opexp}[3]{\ensuremath{\mathcal{D}_{#1}^{#3}(#2)}}
\newcommand{\conj}[2]{\ensuremath{{#1}^{#2}}}

\newcommand{\defeq}{:=}

\DeclareMathOperator{\diag}{diag}

\renewcommand{\vec}[1]{\ensuremath{\bm{#1}}}
\newcommand{\mat}[1]{\ensuremath{\bm{#1}}}
\newcommand{\Mat}[1]{\ensuremath{\bm{#1}}}

\newcommand{\genNorm}[2]{\ensuremath{\mathcal{N}_{#1}(#2)}}

\renewcommand{\a}{\mathbf a}

\newcommand{\f}{\mathbf f}

\newcommand{\Q}{\mathbf Q}

\newcommand{\vecbeta}{\ensuremath{\boldsymbol{\beta}}}

\newcommand{\intSkewRS}[1]{\ensuremath{\mathrm{I}\mathrm{SRS}[#1]}}

\newcommand{\intLinRS}[1]{\ensuremath{\mathrm{I}\mathrm{LRS}[#1]}}
\newcommand{\liftedIntLinRS}[1]{\ensuremath{\mathrm{LILRS}[#1]}}

\newcommand{\SumSubspaceDist}{d_{\ensuremath{\Sigma}S}}

\newcommand{\myspace}[1]{\mathcal{#1}}
\newcommand{\Rowspace}[1]{\ensuremath{{\left\langle #1 \right\rangle}_{q}}}

\newcommand{\RowspaceHuge}[1]{\ensuremath{\scaleleftright[3ex]{\Biggl\langle}{\!\!\!#1\!\!\!}{\Biggr\rangle}\!\!\!\!\raisebox{-21pt}{\scriptsize{q}}}}

\newcommand{\Grassm}[1]{\myspace{G}_q(#1)}

\newcommand{\ProjspaceAny}[1]{\myspace{P}_q(#1)}

\newcommand{\oh}[1]{\bnd{O}{#1}}
\newcommand{\softoh}[1]{\bnd{\widetilde{O}}{#1}}
\newcommand{\bnd}[2]{\ensuremath{#1\mathopen{}\left(#2\right)\mathclose{}}}

\newcommand{\nTransmit}{\ensuremath{n_t}}
\newcommand{\nReceive}{\ensuremath{n_r}}
\newcommand{\insertions}{\ensuremath{\gamma}}
\newcommand{\deletions}{\ensuremath{\delta}}

\newcommand{\pe}{\ensuremath{\alpha}}

\newcommand{\degConstraint}{\ensuremath{D}}
\newcommand{\intOrder}{\ensuremath{s}}

\newcommand{\shots}{\ensuremath{\ell}}
\newcommand{\delOp}[1]{\ensuremath{\mathcal{H}_{#1}}}

\newcommand{\nTransmitShot}[1]{\ensuremath{n_t^{(#1)}}}

\newcommand{\RFmat}{\ensuremath{\Q_R}}
\newcommand{\RFvec}{\ensuremath{\f_R}}

\begin{acronym}
 \acro{SRS}{skew Reed--Solomon}
 \acro{ISRS}{interleaved skew Reed--Solomon}
 \acro{LRS}{linearized Reed--Solomon}
 \acro{LLRS}{lifted linearized Reed--Solomon}
 \acro{ILRS}{interleaved linearized Reed--Solomon}
 \acro{LILRS}{lifted interleaved linearized Reed--Solomon}
\end{acronym}

\newcolumntype{M}[1]{>{\centering\arraybackslash}m{#1}}

\pgfplotsset{
compat=1.8,
mystyle/.style={
    scale only axis,
    width=0.8\columnwidth,
    height=0.6\columnwidth,
    label style={inner sep=0, font=\normalsize}, 
    tick label style={font=\scriptsize},
    legend style={font=\scriptsize},
    mark size=3,
    major grid style={dashed},
    line width=0.8pt,
    axis line style = thin}
}

\IEEEoverridecommandlockouts

\begin{document}

\title{Decoding of Interleaved Linearized Reed--Solomon Codes with Applications to Network Coding}

\author{
	\IEEEauthorblockN{Hannes Bartz}
	\IEEEauthorblockA{Institute of Communications and Navigation \\ German Aerospace Center (DLR), Germany\\
	\texttt{hannes.bartz@dlr.de}}
	
	\and

	\IEEEauthorblockN{Sven Puchinger}
	% \IEEEauthorblockA{Department of Applied Mathematics and Computer Science \\ Technical University of Denmark (DTU), Denmark\\
	% \texttt{svepu@dtu.dk}}
	\IEEEauthorblockA{Institute for Communications Engineering \\ Technical University of Munich (TUM), Germany\\
	\texttt{sven.puchinger@tum.de}}
	\thanks{Sven Puchinger has received funding from the European Union’s Horizon 2020 research and innovation program under the Marie Sklodowska-Curie grant agreement no. 713683.}
}
\maketitle

\begin{abstract}
Recently, Mart{\'\i}nez-Pe{\~n}as and Kschischang (IEEE Trans.\ Inf.\ Theory, 2019) showed that lifted linearized Reed--Solomon codes are suitable codes for error control in multi-shot network coding.
We show how to construct and decode lifted \emph{interleaved} linearized Reed--Solomon codes.
Compared to the construction by Mart{\'\i}nez-Pe{\~n}as--Kschischang, interleaving allows to increase the decoding region significantly (especially w.r.t.\ the number of insertions) and decreases the overhead due to the lifting (i.e., increases the code rate), at the cost of an increased packet size.
The proposed decoder is a list decoder that can also be interpreted as a probabilistic unique decoder. 
Although our best upper bound on the list size is exponential, we present a heuristic argument and simulation results that indicate that the list size is in fact one for most channel realizations up to the maximal decoding radius.
\end{abstract}


%----------------------------------------------------------------------------
% Introduction
%----------------------------------------------------------------------------
\section{Introduction}\label{sec:introduction}

Network coding~\cite{ahlswede2000network} is a powerful approach to achieve the capacity of multicast networks. 
Unlike the classical routing schemes, network coding allows to mix (e.g. linearly combine) incoming packets at intermediate nodes. 
Kötter and Kschischang proposed codes in the subpsace metric as a suitable tool for error correction in (random) linear network coding~\cite{koetter2008coding}.

The sum-rank metric is a hybrid between the Hamming and the rank metric.
Already in \cite{nobrega2009multishot}, codes in this metric (called ``extended rank metric'' therein) were proposed for error correction in multishot network coding. 
Since then, there have been many works on code constructions and efficient decoding algorithms for codes in the sum-rank metric, including \cite{wachter2011partial,wachter2012rank,wachter2015convolutional,napp2017mrd,napp2018faster,martinez2018skew,boucher2019algorithm,martinez2019reliable,caruso2019residues,bartz2020fast,martinezpenas2020sumrank,byrne2020fundamental}.
The first general class of codes attaining the Singleton bound in the sum-rank metric is called \ac{LRS} codes \cite{martinez2018skew}, which can be seen as a mix of Reed--Solomon codes (Hamming metric) and Gabidulin codes (rank metric) \cite{Delsarte_1978,Gabidulin_TheoryOfCodes_1985,Roth_RankCodes_1991}.
It was shown in \cite{martinez2019reliable} that \emph{lifted} \ac{LRS} codes provide reliable and secure coding schemes for non-coherent network coding under an adversarial model.

An $\intOrder$-interleaved code is a direct sum of $\intOrder$ codes of the same length (called constituent codes).
This means that if the constituent codes are over $\Fq$, then the interleaved code can be viewed as a (not necessarily linear) code over $\Fqs$.
In the Hamming and rank metric, there are various decoders that can significantly increase the decoding radius of a constitutent code by collaboratively decoding in an interleaved variant thereof. 
Such decoders are known in the Hamming metric for Reed--Solomon \cite{krachkovsky1997decoding,bleichenbacher2003decoding,coppersmith2003reconstructing,parvaresh2004multivariate,brown2004probabilistic,parvaresh2007algebraic,schmidt2007enhancing,schmidt2009collaborative,cohn2013approximate,nielsen2013generalised,wachterzeh2014decoding,puchinger2017irs,yu2018simultaneous} and in general algebraic geometry codes \cite{brown2005improved,kampf2014bounds,puchinger2019improved}, and in the rank metric for Gabidulin codes \cite{loidreau2006decoding,sidorenko2010decoding,sidorenko2011skew,sidorenko2014fast,wachter2014list,puchinger2017row,puchinger2017alekhnovich,bartz2020fast}.
All of these decoders have in common that they are either list decoders with exponential worst-case and small average-case list size, or probabilistic unique decoders that fail with a very small probability.

Interleaving was suggested in \cite{silva2008rank} as a method to decrease the overhead in lifted Gabidulin codes for error correction in non-coherent (single-shot) network coding, at the cost of a larger packet size while preserving a low decoding complexity. It was later shown \cite{sidorenko2011skew,wachter2014list,bartz2018efficient} that it can also increase the error-correction capability of the code using suitable decoders for interleaved Gabidulin codes.

\subsection{Our Contributions}

In this paper, we define lifted interleaved linearized Reed--Solomon codes and propose a novel interpolation-based list decoder that is based on the list decoder by Wachter-Zeh and Zeh \cite{wachter2014list} for interleaved Gabidulin codes.
We derive a decoding region for the codes in the sum-subspace metric,
analyze the complexity of the decoder,
give an exponential upper bound on the list size,
and give heuristic arguments and numerical evidence that the output list size is with overwhelming probability equal to one for random realizations of the multishot operator channel that stay within the decoding region.

Compared to \cite{martinez2019reliable}, we
decrease the relative overhead introduced by lifting (or equivalently, increase the rate for the same code length and block size) and at the same time extend the decoding region significantly, especially the resilience against insertions in the multishot operator channel.
These advantages come at the cost of 
a larger packet size of the packets within the network and
a supposedly small failure probability.
It is work in progress to derive a formal bound on the failure probability.

Moreover, for the case $\intOrder=1$ (no interleaving), our algorithm does not require the assumption from~\cite[Sec.~V.H]{martinez2019reliable} that $\nReceive\leq\nTransmit$, where $\nTransmit$ and $\nReceive$ denotes the sum of the dimensions of the transmitted and received subspaces, respectively.
Hence, the proposed decoder works in cases in which \cite{martinez2019reliable} does not work.

%----------------------------------------------------------------------------
% Preliminaries 
%----------------------------------------------------------------------------
\section{Preliminaries}\label{sec:preliminaries}

Let $\Fq$ be a finite field of order $q$ and denote by $\Fqm$ the extension field of $\Fq$ of degree $m$ with primitive element $\pe$.
The multiplicative group $\Fqm\setminus\{0\}$ of $\Fqm$ is denoted by $\Fqm^*$.
Matrices and vectors are denoted by bold uppercase and lowercase letters like $\mat{A}$ and $\vec{a}$, respectively.
Under a fixed basis of $\Fqm$ over $\Fq$ any element $a\in\Fqm$ can be represented by a corresponding vector $\vec{a}\in\Fq^m$.
For $\mat{A}\in\Fqm^{M\times N}$ we denote by $\Rowspace{\mat{A}}$ the $\Fq$-linear rowspace of the matrix $\mat{A}_q\in\Fq^{M\times Nm}$ obtained by row-wise expanding the elements in $\mat{A}$ over~$\Fq$.

\subsection{Skew Polynomials}
\emph{Skew polynomials} are non-commutative polynomials that were introduced by Ore~\cite{ore1933theory}.
Let $\aut:\Fqm\to\Fqm$ be a finite field automorphism.
A \emph{skew polynomial} is a polynomial of the form 
\begin{equation}
 \textstyle f(x)=\sum_{i}f_i x^i
\end{equation}
with a finite number of coefficients $f_i\in\Fqm$ being nonzero.
The degree $\deg(f)$ of a skew polynomial $f$ is defined as $\max\{i:f_i\neq 0\}$ if $f\neq0$ and $-\infty$ otherwise.

The set of skew polynomials with coefficients in $\Fqm$ together with ordinary polynomial addition and the multiplication rule
\begin{equation}
  xa=\aut(a)x,\qquad a\in\Fqm
\end{equation}
forms a non-commutative ring denoted by $\SkewPolyringZeroDer$.
The set of skew polynomials in $\SkewPolyringZeroDer$ of degree less than $k$ is denoted by $\SkewPolyringZeroDer_{<k}$.
For an element $a\in\Fqm$ the generalized power function is defined as (see~\cite{lam1988vandermonde})
\begin{equation}
  \genNorm{i}{a} := \aut^{i-1}(a)\aut^{i-2}(a)\dots\aut(a)a.
\end{equation}
For any $a,b\in\Fqm$ we define the operator $\op{a}{b} := \aut(b)a$ and $\opexp{a}{b}{i} := \aut^i(b)\aut^{i-1}(a)\dots\aut(a)a=\aut^i(b)\genNorm{i}{a}$.
The \emph{generalized operator evaluation} of a skew polynomial $f\in\SkewPolyringZeroDer$ at an element $b$ wrt. $a$, where $a,b\in\Fqm$, is defined as
\begin{equation}
  \textstyle \opev{f}{b}{a}=\sum_{i}f_i\opexp{a}{b}{i}.
\end{equation}

\begin{lemma}[Product Rule~\cite{martinez2019private}]\label{lem:opev_product_rule}
For two skew polynomials $f,g\in\SkewPolyringZeroDer$ and elements $a,b\in\Fqm$ the generalized operator evaluation of the product $f\cdot g$ at $b$ w.r.t $a$ satisfies 
\begin{equation}
	\opev{(f\cdot g)}{b}{a} = \opev{f}{\opev{g}{b}{a}}{a}.
\end{equation}
\end{lemma}

\subsection{Conjugacy Classes}

Two elements $a,b\in\Fqm$ are called \emph{conjugates} if there exists an element $c\in\Fqm^*$ s.t. $\conj{a}{c}\defeq\aut(c)ac^{-1}=b$.
The set 
\begin{equation}
  \set{C}(a)\defeq\left\{a^c:c\in\Fqm^*\right\}
\end{equation}
is called \emph{conjugacy class} of $a$.
A finite field $\Fqm$ has at most $\shots\leq q-1$ distinct conjugacy classes.
For $\shots\leq q-1$ the elements $1,\pe,\pe^2,\dots,\pe^{\shots-2}$ are representatives of all (nontrivial) disjoint conjugacy classes of $\Fqm$.

\begin{proposition}[Number of Roots~\cite{caruso2019residues}]\label{prop:number_of_roots}
Let $\beta_1^{(i)},\dots,\beta_{n_i}^{(i)}$ be elements from $\Fqm$ and let $a_1,\dots,a_\shots$ be representatives be from conjugacy classes of $\Fqm$ for all $i=1,\dots,\shots$.
Then for any nonzero $f\in\SkewPolyringZeroDer$ satisfying 
\begin{equation}
 	\opev{f}{\beta_j^{(i)}}{a_i}=0,\forall i=1,\dots,\shots,j=1,\dots,n_i
\end{equation}
we have that
$
\deg(f)\leq\sum_{i=1}^{\shots}n_i
$
where equality holds iff the $\beta_1^{(i)},\dots,\beta_{n_i}^{(i)}$ are $\Fq$-linearly independent for each $i=1,\dots,\shots$.
\end{proposition}

\subsection{Multishot Network Coding}

As a channel model we consider the \emph{multishot operator channel} from~\cite{nobrega2009multishot} which consists of multiple independent channel uses of the operator channel from~\cite{koetter2008coding}. 
Let $\ProjspaceAny{N_i}$ denote the set of all subspaces of $\Fq^{N_i}$.
For $N=N_1+N_2+\dots+N_\shots$ we define the $\shots$-fold Cartesian product
\begin{equation}
\ProjspaceAny{N} := \textstyle \prod_{i=1}^{\shots}\ProjspaceAny{N_i}=\ProjspaceAny{N_1}\times\dots\times\ProjspaceAny{N_\shots}.
\end{equation}
The Grassmannian, i.e. the set of all subspaces of dimension $l$ in $\ProjspaceAny{N_i}$, is denoted by $\Grassm{N_i,l}$.

We now consider $\shots$ independent channel uses of the operator channel~\cite{koetter2008coding}.
We consider \emph{sum-constant-dimension} codes, i.e. codes that inject the same number of (linearly independent) packets $\nTransmitShot{i}$ in a given shot.
In the $i$-th shot, the operator channel takes a subspace $\myspace{V}_i\in\Grassm{N_i,\nTransmitShot{i}}$ and returns a subspace
\begin{equation}\label{eq:def:multishot_op_channel}
\myspace{U}_i=\delOp{\nTransmitShot{i}-\deletions^{(i)}}(\myspace{V}_i)\oplus\myspace{E}_i
\end{equation}
where $\delOp{\nTransmitShot{i}-\deletions^{(i)}}$ returns a random $(\nTransmitShot{i}-\deletions^{(i)})$-dimensional subspace of $\myspace{V}_i$ and $\myspace{E}_i$ is an error subspace of dimension $\insertions^{(i)}=\dim(\myspace{E}_i)$ with $\myspace{V}_i\cap\myspace{E}_i=\vec{0}$.
Hence, each received subspace $\myspace{U}_i$ has dimension $\nReceive^{(i)}\defeq\dim(\myspace{U}_i)=\nTransmitShot{i}+\insertions^{(i)}-\deletions^{(i)}$.
The overall transmitted/received words are tuples of subspaces
\begin{align}
	\vec{\myspace{V}}&=\left(\myspace{V}_1,\myspace{V}_2,\dots,\myspace{V}_\shots\right)
	\in \textstyle\prod_{i=1}^{\shots}\Grassm{N_i,\nTransmitShot{i}}, \\
 \vec{\myspace{U}}&=\left(\myspace{U}_1,\myspace{U}_2,\dots,\myspace{U}_\shots\right)
	\in\textstyle\prod_{i=1}^{\shots}\Grassm{N_i,\nReceive^{(i)}},
\end{align}
where we define $\nTransmit=\sum_{i=1}^{\shots}\nTransmitShot{i}$ and $\nReceive=\sum_{i=1}^{\shots}\nReceive^{(i)}$.
After $\shots$ channel uses (or \emph{shots}) we have that
\begin{equation}\label{eq:sum_dimensions}
 \nReceive=\textstyle\sum_{i=1}^{\shots}\nReceive^{(i)}=\sum_{i=1}^{\shots}\nTransmitShot{i}+\insertions^{(i)}-\deletions^{(i)}.
\end{equation}
Setting $\insertions=\sum_{i=1}^{\shots}\insertions^{(i)}$, $\deletions=\sum_{i=1}^{\shots}\deletions^{(i)}$ we may write~\eqref{eq:sum_dimensions} as
\begin{equation}
	\nReceive = \nTransmit+\insertions-\deletions.
\end{equation}

\begin{definition}[Sum-Subspace Distance~\cite{nobrega2009multishot}]
 Given $\vec{\myspace{U}}=(\myspace{U}_1,\myspace{U}_2,\dots,\myspace{U}_\shots)$ and $\vec{\myspace{V}}=(\myspace{V}_1,\myspace{V}_2,\dots,\myspace{V}_\shots)\in\ProjspaceAny{N}$ the sum-subspace distance between $\vec{\myspace{U}}$ and $\vec{\myspace{V}}$ is defined as
 \begin{align}\label{eq:def_sum_subspace_dist}
 	\SumSubspaceDist(\vec{\myspace{U}},\vec{\myspace{V}})&\defeq \textstyle\sum_{i=1}^{\shots}\left(\dim(\myspace{U}_i+\myspace{V}_i)-\dim(\myspace{U}_i\cap\myspace{V}_i)\right).
 \end{align}
\end{definition}

Similar to~\cite{bartz2020fast} we define the $(\insertions,\deletions)$ reachability for multiple independent operator channel uses.

\begin{definition}[$(\insertions,\deletions)$ Reachability]\label{def:ins_del_reachability}
 Given two tuples of subspaces $\vec{\myspace{U}}=(\myspace{U}_1,\myspace{U}_2,\dots,\myspace{U}_\shots)$ and $\vec{\myspace{V}}=(\myspace{V}_1,\myspace{V}_2,\dots,\myspace{V}_\shots)\in\ProjspaceAny{N}$ we say that $\vec{\myspace{V}}$ is $(\insertions,\deletions)$-reachable from $\vec{\myspace{U}}$ if there exists a realization of the multishot operator channel~\eqref{eq:def:multishot_op_channel} with $\insertions=\sum_{i=1}^{\shots}\insertions^{(i)}$ insertions and $\deletions=\sum_{i=1}^{\shots}\deletions^{(i)}$ deletions that transforms the input $\vec{\myspace{V}}$ to the output $\vec{\myspace{U}}$.
\end{definition}

\begin{proposition}
 Consider $\vec{\myspace{U}}=(\myspace{U}_1,\myspace{U}_2,\dots,\myspace{U}_\shots)$ and $\vec{\myspace{V}}=(\myspace{V}_1,\myspace{V}_2,\dots,\myspace{V}_\shots)\in\ProjspaceAny{N}$.
 If $\vec{\myspace{V}}$ is $(\insertions,\deletions)$-reachable from $\vec{\myspace{U}}$, then we have that $\SumSubspaceDist(\vec{\myspace{U}},\vec{\myspace{V}})=\insertions+\deletions$.
\end{proposition}

%----------------------------------------------------------------------------
% Lifted Interleaved Linearized Reed--Solomon Codes
%----------------------------------------------------------------------------
\section{Lifted Interleaved Linearized Reed--Solomon Codes}\label{sec:LILRS}

In this section we consider \ac{LILRS} codes for multiple transmissions over the operator channel.
We generalize the ideas from~\cite{martinez2019reliable} to obtain multishot subspace codes by \emph{lifting} \ac{ILRS} codes.

\begin{definition}[Lifted Interleaved Linearized RS Code]\label{def:LILRS}
Let $\a=(a_1,a_2,\dots,a_\shots)$ be a vector containing representatives from different conjugacy classes of $\Fqm$.
Let the vectors $\vecbeta^{(i)}=(\beta_1^{(i)},\beta_2^{(i)},\dots,\beta_{\nTransmitShot{i}}^{(i)})\in\Fqm^{\nTransmitShot{i}}$ contain $\Fq$-linearly independent elements from $\Fqm$ for all $i=1,\dots,\shots$ and define $\vecbeta=\left(\vecbeta^{(1)}\mid\vecbeta^{(2)}\mid\dots\mid\vecbeta^{(\shots)}\right)\in\Fqm^{\nTransmit}$.
A lifted $\intOrder$-interleaved linearized Reed--Solomon (LILRS) code $\liftedIntLinRS{\vecbeta,\a,\shots,\intOrder;\nTransmit,k}$ of subspace dimension $\nTransmit=\nTransmit^{(1)}+\nTransmit^{(2)}+\dots+\nTransmit^{(\shots)}$ and dimension $k\leq\nTransmit$ is defined as

\begin{align*}
\Big\{\vec{\myspace{V}}(\f) \defeq \left(\myspace{V}_1(\f), \dots, \myspace{V}_\shots(\f)\right) \, : \, \f \in \SkewPolyringZeroDer_{<k}^\intOrder\Big\} \\
\subseteq \textstyle \prod_{i=1}^{\shots} \Grassm{N_i,\nTransmitShot{i}}
\end{align*}
where $N_i=\nTransmitShot{i}+\intOrder m$, and, for $\f = (f^{(1)}, \dots, f^{(\intOrder)})$, we have
\begin{equation*}
\myspace{V}_i(\f)\!\defeq\!
	\RowspaceHuge{
	\begin{pmatrix}
	 \beta_1^{(i)} & \opev{f^{(1)}}{\beta_1^{(i)}}{a_i} & \!\!\dots\!\! & \opev{f^{(\intOrder)}}{\beta_1^{(i)}}{a_i}\!\!
	 \\[-2pt]
	 \!\vdots\! & \!\vdots\! & \!\!\ddots\!\! & \!\vdots\!
	 \\ 
	 \beta_{\nTransmitShot{i}}^{(i)} & \opev{f^{(1)}}{\beta_{\nTransmitShot{i}}^{(i)}}{a_i} & \!\!\dots\!\! & \opev{f^{(\intOrder)}}{\beta_{\nTransmitShot{i}}^{(i)}}{a_i}
	\end{pmatrix}
	}\,.
\end{equation*}
\end{definition}

The code rate of an LILRS code $\liftedIntLinRS{\vecbeta,\a,\shots,\intOrder;\nTransmit,k}$ is defined as
\begin{equation}\label{def:code_rate_LILRS}
 R
 =\frac{\log_q(|\liftedIntLinRS{\vecbeta,\a,\shots,\intOrder;\nTransmit,k}|)}{\sum_{i=1}^{\shots}\nTransmitShot{i}N_i}
 =\frac{\intOrder mk}{\sum_{i=1}^{\shots}\nTransmitShot{i}(\nTransmitShot{i}+\intOrder m)}.
\end{equation}
Note, that there exist other definitions of the code rate for multishot codes, which are discussed in~\cite[Section~IV.A]{nobrega2009multishot}.

The definition of~\ac{LILRS} codes generalizes several code families.  
For $\intOrder=1$ we obtain the lifted linearized Reed--Solomon codes from~\cite[Section~V.III]{martinez2019reliable}. 
For $\shots=1$ we obtain lifted interleaved Gabidulin codes as considered in e.g.~\cite{wachter2014list, bartz2018efficient} with Kötter--Kschischang codes~\cite{koetter2008coding} as special case for $\intOrder=1$.
Without lifting we obtain interleaved linearized Reed--Solomon codes with linearized Reed--Solomon codes~\cite{martinez2018skew} as special case for $\intOrder=1$.

\begin{proposition}
 The minimum sum-subspace distance of a \ac{LILRS} code $\liftedIntLinRS{\vecbeta,\a,\shots,\intOrder;\nTransmit,k}$ as in Definition~\ref{def:LILRS} is
 \begin{equation}
 	\SumSubspaceDist\left(\liftedIntLinRS{\vecbeta,\a,\shots,\intOrder;\nTransmit,k}\right)=2\left(\nTransmit-k+1\right).
 \end{equation}
\end{proposition}

%----------------------------------------------------------------------------
% An Interpolation-Based Decoding Approach
%----------------------------------------------------------------------------
\section{An Interpolation-Based Decoding Approach}\label{sec:decodingLILRS}

We now derive an interpolation-based decoding approach for~\ac{LILRS} codes.
The decoding principle consists of an interpolation step and a root-finding step.
In~\cite{martinez2019reliable}, (lifted) linearized Reed--Solomon codes are decoded using the isometry between the sum-rank and the skew metric.
In this paper we consider an interpolation-based decoding scheme in the generalized operator evaluation domain without the need for casting the decoding problem to the skew metric.
The new decoder is a generalization of \cite{wachter2014list} (interleaved Gabidulin codes in the rank metric) and \cite{bartz2018efficient} (lifted interleaved Gabidulin codes in the subspace metric).

\subsection{Interpolation Step}

Suppose we transmit the tuple of subspaces 
\begin{equation}
 \vec{\myspace{V}}(\f) = \left(\myspace{V}_1(\f), \dots, \myspace{V}_\shots(\f)\right)\in\textstyle\prod_{i=1}^{\shots} \Grassm{N_i,\nTransmitShot{i}}
\end{equation}
over the multishot operator channel with overall $\insertions$ insertions and $\deletions$ deletions and receive the subspaces
\begin{equation}
 \vec{\myspace{U}} = \left(\myspace{U}_1, \dots, \myspace{U}_\shots\right)\in \textstyle\prod_{i=1}^{\shots} \Grassm{N_i,\nReceive^{(i)}},
\end{equation}
where, for all $i=1,\dots,\shots$,
\begin{equation}
	\myspace{U}_i = 	
	\RowspaceHuge{
	\begin{pmatrix}
	 \xi_1^{(i)} & u_{1}^{(1,i)} & u_{1}^{(2,i)} & \dots & u_{1}^{(\intOrder,i)}
	 \\[-2pt] 
	 \vdots & \vdots & \vdots & \ddots & \vdots
	 \\ 
	 \xi_{\nReceive^{(i)}}^{(i)} & u_{\nReceive^{(i)}}^{(1,i)} & u_{\nReceive^{(i)}}^{(2,i)} & \dots & u_{\nReceive^{(i)}}^{(\intOrder,i)}
	\end{pmatrix}
	}\,.
\end{equation}
\vspace{-0.7cm}
\begin{remark}
 In contrast to~\cite[Section~V.III]{martinez2019reliable} we do not need the assumption that the dimension(s) of the transmitted subspace(s) equals the dimension(s) of the received subspace(s).
\end{remark}
For a multivariate skew polynomial of the form
\begin{equation}\label{eq:mult_var_skew_poly}
	Q(x, y_1,\dots, y_\intOrder)=Q_0(x)+Q_1(x)y_1+\dots+Q_\intOrder(x)y_\intOrder
\end{equation}
where $Q_l(x)\in\SkewPolyringZeroDer$ for all $l\in\intervallincl{0}{\intOrder}$ define the $\nReceive$ evaluations $\mathscr{E}_j^{(i)}$ for $j=1,\dots,\nReceive^{(i)}$ and $i=1,\dots,\shots$ as
\begin{equation}\label{eq:defDiGenOpLILRS}
	\mathscr{E}_j^{(i)}(Q)\defeq \opev{Q_0}{\xi_j^{(i)}}{a_i}+\textstyle\sum_{l=1}^{\intOrder}\opev{Q_l}{u_j^{(l,i)}}{a_i}.
\end{equation}
The $\vec{w}$-weighted degree of a multivariate skew polynomial $Q$ as in~\eqref{eq:mult_var_skew_poly} is defined as $\deg_{\vec{w}}(Q)\!=\!\max_j\{\deg(Q_j)\!+\!w_j\}$.
Now consider the following interpolation problem in $\SkewPolyringZeroDer$. 

\begin{problem}[Generalized Operator Interpolation Problem]\label{prob:skewIntProblemGenOpLILRS}
 Given the integers $\degConstraint,\intOrder\in\mathbb{Z}_+$, a set  
 \begin{equation}
  \set{E}=\left\{\mathscr{E}_j^{(i)}:i=1,\dots,\shots, j=1,\dots,n_i\right\}
 \end{equation}
 containing the generalized operator evaluation maps defined in~\eqref{eq:defDiGenOpLILRS} and a vector $\vec{w}=(0,k-1,\dots,k-1)\in\ZZ_+^{\intOrder+1}$, find a nonzero polynomial of the form 
 \begin{equation}
 	Q(x, y_1,\dots, y_\intOrder)=Q_0(x)+Q_1(x)y_1+\dots+Q_\intOrder(x)y_\intOrder
 \end{equation}
 with $Q_l(x)\in\SkewPolyringZeroDer$ for all $l\in\intervallincl{0}{\intOrder}$ that satisfies:
 \begin{enumerate}
  \item $\mathscr{E}_j^{(i)}(Q)=0, \qquad\forall i=1,\dots,\shots$, $j=1,\dots,\nReceive^{(i)}$,
  \item $\deg_{\vec{w}}(Q(x,y_1,\dots,y_\intOrder))<\degConstraint$.
 \end{enumerate}
\end{problem}
A solution of Problem~\ref{prob:skewIntProblemGenOpLILRS} can be found by calling the skew Kötter interpolation~\cite{liu2014kotter} with evaluation maps $\mathscr{E}_j^{(i)}$ as defined in~\eqref{eq:defDiGenOpLILRS} requiring $\oh{\intOrder^2 n^2}$ operations in $\Fqm$.
\begin{lemma}[Existence of Solution]
 A nonzero solution of Problem~\ref{prob:skewIntProblemGenOpLILRS} exists if
$\degConstraint=\big\lceil\tfrac{\nReceive+\intOrder(k-1)+1}{\intOrder+1}\big\rceil$.
\end{lemma}

\begin{IEEEproof}
 Problem~\ref{prob:skewIntProblemGenOpLILRS} corresponds to a system of $\nReceive$ $\Fqm$-linear equations in $\degConstraint(\intOrder+1)-\intOrder(k-1)$ unknowns which has a nonzero solution the number of equations is less than the number of unknowns, i.e. if
 \begin{equation}
  \nReceive<\degConstraint(\intOrder+1)-\intOrder(k-1)\label{eq:existenceCondLILRS}
  % \\
  \quad\Longleftrightarrow\quad
  \degConstraint\geq\tfrac{\nReceive+\intOrder(k-1)+1}{\intOrder+1}.
 \end{equation} 
\end{IEEEproof}

\subsection{Root-Finding Step}

The goal of the root-finding step is to recover the message polynomials $f^{(1)},\dots,f^{(\intOrder)}\in\SkewPolyringZeroDer_{<k}$ from the multivariate polynomial constructed in the interpolation step. 
Therefore, we need the following results.

\begin{lemma}[Roots of Polynomial]\label{lem:decConditionLILRS}
 Let
 \begin{equation}
 	P(x)\defeq Q_0(x)+Q_1(x)f^{(1)}(x)+\dots+Q_\intOrder(x)f^{(\intOrder)}(x).
 \end{equation}
 Then there exist elements $\zeta_1^{(i)},\dots,\zeta_{\nTransmitShot{i}-\deletions^{(i)}}^{(i)}$ in $\Fqm$ that are $\Fq$-linearly independent for each $i=1,\dots,\shots$ such that
 \begin{equation}
 	\opev{P}{\zeta_j^{(i)}}{a_i}=0
 \end{equation}
 for all $i=1,\dots,\shots$ and $j=1,\dots,\nTransmitShot{i}-\deletions^{(i)}$.
\end{lemma}

\begin{IEEEproof}
 In each shot the noncorrupted intersection space has dimension $\dim(\myspace{U}_i\cap\myspace{V}_i)=\nTransmitShot{i}-\deletions^{(i)}$ for all $i=1,\dots,\shots$.
 A basis for each intersection space $\myspace{U}_i\cap\myspace{V}_i$ can be represented as
 \begin{equation}
    \left\{\left(\zeta_j^{(i)},\opev{f^{(1)}}{\zeta_j^{(i)}}{a_i},\!\dots,\opev{f^{(\intOrder)}}{\zeta_j^{(i)}}{a_i}\right)\!:\!j\!\in\!\intervallincl{1}{\nTransmitShot{i}\!-\!\deletions^{(i)}}\right\}
 \end{equation}
 where $\zeta_1^{(i)},\dots,\zeta_{\nTransmitShot{i}-\deletions^{(i)}}^{(i)}$ are $(\nTransmitShot{i}-\deletions^{(i)})$ $\Fq$-linearly independent elements from $\Fqm$ for all $i=1,\dots,\shots$.
 Since each intersection space $\myspace{U}_i\cap\myspace{V}_i$ is a subspace of the received space $\myspace{U}_i$ we have that
 \begin{equation}\label{eq:decConditionLILRS}
 	\opev{P}{\zeta_j^{(i)}}{a_i}\defeq\opev{Q_0}{\zeta_j^{(i)}}{a_i}+\sum_{l=1}^{\intOrder}\opev{Q_l}{\opev{f^{(l)}}{\zeta_j^{(i)}}{a_i}}{a_i}
 	=0
 \end{equation}
 for all $i=1,\dots,\shots, j=1,\dots,\nTransmitShot{i}-\deletions^{(i)}$.
\end{IEEEproof} 

\begin{theorem}[Decoding Region] 
 Let $\vec{\myspace{U}} \in \prod_{i=1}^{\shots}\Grassm{N_i,\nReceive^{(i)}}$ be the tuple containing the received subspaces and let $Q(x,y_1,\dots,y_\intOrder)\neq0$ fulfill the constraints in Problem~\ref{prob:skewIntProblemGenOpLILRS}. 
Then for all codewords $\vec{\myspace{V}}(\f)\in\liftedIntLinRS{\vecbeta,\a,\shots,\intOrder;\nTransmit,k}$ that are $(\insertions,\deletions)$-reachable from $\vec{\myspace{U}}$, where $\insertions$ and $\deletions$ satisfy 
 \begin{equation}\label{eq:listDecRegion}
  \insertions+\intOrder \deletions<\intOrder(\nTransmit-k+1),
 \end{equation}
 we have that
 \begin{equation}\label{eq:rootFindingEquation}
 	P(x)\!=\!Q_0(x)+Q_1(x)f^{(1)}(x)+\!\dots\!+Q_\intOrder(x)f^{(\intOrder)}(x)\!=\!0.
 \end{equation}
\end{theorem}

\begin{IEEEproof}
 By Lemma~\ref{lem:decConditionLILRS} there exist elements $\zeta_1^{(i)},\dots,\zeta_{\nTransmitShot{i}-\deletions^{(i)}}^{(i)}$ in $\Fqm$ that are $\Fq$-linearly independent for each $i=1,\dots,\shots$ such that
 \begin{equation}
 	\opev{P}{\zeta_j^{(i)}}{a_i}=0
 \end{equation}
 for all $i=1,\dots,\shots$ and $j=1,\dots,\nTransmitShot{i}-\deletions^{(i)}$.
 By choosing 
 \begin{equation}\label{eq:decDegreeConstraintLILRS}
 	\degConstraint\leq \nTransmit-\deletions
 \end{equation}
 the degree of $P(x)$ exceeds the degree bound from Proposition~\ref{prop:number_of_roots} which is possible only if $P(x)=0$.
 Combining~\eqref{eq:existenceCondLILRS} and~\eqref{eq:decDegreeConstraintLILRS} we get
 \begin{align*}
  \nReceive+\intOrder(k-1)&<\degConstraint(\intOrder+1)\leq(\intOrder+1)(\nTransmit-\deletions)
  \\\Longleftrightarrow\qquad
  \insertions+\intOrder\deletions&<\intOrder(\nTransmit-k+1).
 \end{align*}
\end{IEEEproof}

The decoding region in~\eqref{eq:listDecRegion} shows and improved insertion-correction performance due to interleaving.

In the root-finding step, all polynomials $f^{(1)},\dots,f^{(\intOrder)}\in\SkewPolyringZeroDer_{<k}$ that satisfy~\eqref{eq:rootFindingEquation} need to be found. 
This task can be accomplished by the efficient minimal approximant bases algorithm in~\cite{bartz2020fast} with at most $\softoh{\intOrder^{2.372} \nTransmit^{1.635}}$ operations in $\Fqm$, where $\tilde{O}$ is the soft-$O$ notation, which neglects log factors.

\subsection{List Decoding}

Instead of using only one solution of Problem~\ref{prob:skewIntProblemGenOpLILRS}, we follow the ideas of~\cite{wachter2014list} and use a basis of the $\Fqm$-linear solution space of the interpolation problem in order to derive bounds on the worst-case and average list size.
In~\cite{bartz2017algebraic} it was shown that using a degree-restricted subset of a Gröbner basis for the left $\SkewPolyringZeroDer$-linear interpolation submodule achieves the smallest possible list size.
Let the dimension of the $\Fqm$-linear solution space of Problem~\ref{prob:skewIntProblemGenOpLILRS} be $d_I$ and define the corresponding basis polynomials as
\begin{equation}
  Q_0^{(r)}(x)=\textstyle\sum_{i=0}^{\degConstraint-1}q_{0,i}^{(r)}x^i, \quad
  Q_j^{(r)}(x)=\textstyle\sum_{i=0}^{\degConstraint-k}q_{j,i}^{(r)}x^i
\end{equation}
for all $j=1,\dots,\intOrder$ and $r=1,\dots,d_I$.
Define the matrix
 \begin{equation}
  \mat{Q}_j^{i}=
  \begin{pmatrix}
   \aut^i\left(q_{1,j}^{(1)}\right) & \aut^i\left(q_{2,j}^{(1)}\right) & \dots & \aut^i\left(q_{\intOrder,j}^{(1)}\right)
   \\
   \vdots & \vdots & \ddots & \vdots
   \\ 
   \aut^i\left(q_{1,j}^{(d_I)}\right) & \aut^i\left(q_{2,j}^{(d_I)}\right) & \dots & \aut^i\left(q_{\intOrder,j}^{(d_I)}\right)
  \end{pmatrix}\in\Fqm^{d_I\times s}
 \end{equation}
 and the vectors
 \begin{equation}
  \vec{f}_j^{i}\defeq\left(\aut^i\left(f_j^{(1)}\right),\dots,\aut^i\left(f_j^{(\intOrder)}\right)\right)\in\Fqm^{\intOrder}
 \end{equation}
 and
 \begin{equation}
  \vec{q}_{0,j}^i\defeq\left(\aut^i\left(q_{0,j}^{(1)}\right),\dots,\aut^i\left(q_{0,j}^{(d_I)}\right)\right)\in\Fqm^{d_I}.
 \end{equation}
Defining the root-finding matrix 
 \begin{equation}\label{eq:rootFindingMatrixLILRS}
 \RFmat=
 \begin{pmatrix}
 \Mat{Q}_0^{0}\phantom{^{-}}        &     &  &              \\
 \Mat{Q}_1^{-1}         & \Mat{Q}_0^{-1}  &        &              \\[-3pt]
 \vdots             & \Mat{Q}_1^{-2}  & \ddots &              \\[-3pt]
 \Mat{Q}_{\degConstraint-k}^{-(\degConstraint-k)} & \vdots    & \ddots & \Mat{Q}_{0}^{-(k-1)}       \\[-3pt]
              & \ddots    & \ddots & \Mat{Q}_{1}^{-k\phantom{(-1)}}           \\[-5pt]
                                          &     & \ddots & \vdots          \\
              &           & & \Mat{Q}_{\degConstraint-k}^{-(\degConstraint-1)}
 \end{pmatrix}
\end{equation}
and the vectors
\begin{equation*}
 \RFvec=\left(\vec{f}_0,\dots,\vec{f}_{k-1}^{-(k-1)}\right)^T
 \quad\text{and}\quad
 \vec{q}_{0}=\left(\vec{q}_{0,0},\dots, \vec{q}^{-(\degConstraint-1)}_{0,\degConstraint-1}\right)^T
\end{equation*}
we can write the root-finding system~\eqref{eq:rootFindingEquation} as
\begin{equation}\label{eq:rootFindingSystemFqm}
 \RFmat\cdot\RFvec=-\vec{q}_0.
\end{equation}

In general, the root-finding matrix $\RFmat$ in~\eqref{eq:rootFindingMatrixLILRS} can be rank deficient.
In this case we obtain a \emph{list} of potential message polynomials $f^{(1)},\dots,f^{(\intOrder)}$.
Using the same arguments as in \cite{wachter2014list} on the structure of $\RFmat$ in \eqref{eq:rootFindingMatrixLILRS}, one can derive a lower bound on the rank of $\RFmat$, and thus the following upper bound on the number of solutions of \eqref{eq:rootFindingSystemFqm}.

\begin{lemma}[Worst-Case List Size]
 The root-finding system in~\eqref{eq:rootFindingEquation} has at most $q^{m(k(\intOrder-1))}$ solutions $f^{(1)},\dots,f^{(\intOrder)}\in\SkewPolyringZeroDer_{<k}$.
\end{lemma}
In general, we have that $k\leq\nTransmit$, where $\nTransmit\leq\shots m$.
Hence, for $m\approx\nTransmit/\shots$ we get a worst-case list size of $q^{\frac{\nTransmit}{\shots}(k(\intOrder-1))}$.

\subsection{Probabilistic-Unique Decoding}

We now consider the interpolation-based decoder from Section~\ref{sec:decodingLILRS} as a probabilistic-unique decoder which either returns a unique solution (if the list size is equal to one) or a decoding failure.
In order to get an estimate of the decoding failure probability $P_f$, we use similar assumptions as in~\cite{wachter2014list} to derive a heuristic upper bound.  

\begin{figure}[ht!]
\centering
\definecolor{mycolor1}{rgb}{0.00000,0.44700,0.74100}%
\definecolor{mycolor2}{rgb}{0.85000,0.32500,0.09800}%
\begin{tikzpicture}

\begin{axis}[%
xmin=0,
xmax=13,
xtick={ 0,  1,  2,  3,  4,  5,  6,  7,  8,  9, 10, 11, 12, 13},
xlabel={$\insertions\text{ insertions}$},
xmajorgrids,
ymin=0,
ymax=4,
ytick={0, 1, 2, 3, 4},
ylabel={$\deletions\text{ deletions}$},
ymajorgrids,
legend style={at={(0.99,0.965)},anchor=north east,legend cell align=left,align=left, draw=white!15!black},
mystyle,
height=0.33\columnwidth,
width=0.9\columnwidth,
reverse legend
]

\addplot[const plot, color=gray,  dotted, line width=1.5pt] plot table[row sep=crcr]{%
-2	-5\\
-1	3\\
0.02	3\\
1	3\\
2	3\\
3	2\\
4	2\\
5	2\\
6	2\\
7	1\\
8	1\\
9	1\\
10	1\\
11	0.02\\
12	0.02\\
13	0.02\\
14	0.02\\
15	-1\\
16	-1\\
17	-3\\
};
\addlegendentry{List decoding~\eqref{eq:listDecRegion} $(\intOrder=4)$};

\addplot[const plot, fill=cyan,color=cyan,  pattern=north east lines, pattern color = cyan, solid, line width=1.5pt] plot table[row sep=crcr] {%
-3	-3 \\
-2	3\\
-1	3\\
0.01	2\\
1	2\\
2	2\\
3	2\\
4	1\\
5	1\\
6	1\\
7	1\\
8	0.02\\
9	0.02\\
10	0.02\\
11	0.02\\
12	-3\\
};
\addlegendentry{Probabilistic unique decoding~\eqref{eq:decRegionLILRSprob} $(\intOrder=4)$};

\addplot[const plot, fill = blue, color=blue, pattern=north west lines, pattern color = blue,  dashed, line width=1.5pt] plot table[row sep=crcr] {%
-2	-2\\
-1	3\\
0.02	2\\
1	1\\
2	0.02\\
3	-1\\
4	-2\\
5	-2\\
6	-2\\
7	-2\\
8	-2\\
9	-2\\
10	-2\\
11	-2\\
12	-2\\
13	-2\\
};
\addlegendentry{Martinez-Kschischang \cite{martinez2019reliable} $(\intOrder=1)$};

% add arrows
\node[anchor=east] (A) at (axis cs: 0.01, 1.45){};
\node[anchor=west] (B) at (axis cs: 3.965,1.45){};
\draw [<->, line width=1pt] (A) edge (B);
\node (T) at (axis cs: 2,1.65){\footnotesize{$\intOrder$}};

\node[anchor=east] (C) at (axis cs: 0.01, 0.45){};
\node[anchor=west] (D) at (axis cs: 7.965,0.45){};
\draw [<->, line width=1pt] (C) edge (D);
\node (T) at (axis cs: 4,0.65){\footnotesize{$2\intOrder$}};

% add arrows
\node[anchor=east] (A) at (axis cs: 0.05, 2.45){};
\node[anchor=west] (B) at (axis cs: 2.97,2.45){};
\draw [<->, line width=1pt,color=gray] (A) edge (B);
\node[color=gray] (T) at (axis cs: 1.5,2.65){\footnotesize{$\intOrder-1$}};

% add arrows 
\node[anchor=east] (A) at (axis cs: 4.035, 1.45){};
\node[anchor=west] (B) at (axis cs: 6.97,1.45){};
\draw [<->, line width=1pt,color=gray] (A) edge (B);
\node[color=gray] (T) at (axis cs: 5.5,1.65){\footnotesize{$\intOrder-1$}};

% add arrows 
\node[anchor=east] (A) at (axis cs: 8.035, 0.45){};
\node[anchor=west] (B) at (axis cs: 10.97,0.45){};
\draw [<->, line width=1pt,color=gray] (A) edge (B);
\node[color=gray] (T) at (axis cs: 9.5,0.65){\footnotesize{$\intOrder-1$}};

\end{axis}  

\end{tikzpicture}%
\caption{Decoding region for Martinez-Kschischang~\cite{martinez2019reliable} codes $(\intOrder=1)$ and for decoding of lifted $(\intOrder=4)$-interleaved linearized Reed--Solomon codes.
 The decoding region for insertions increases with the interleaving order $\intOrder$.
}\label{fig:decodingRegion}
\end{figure}
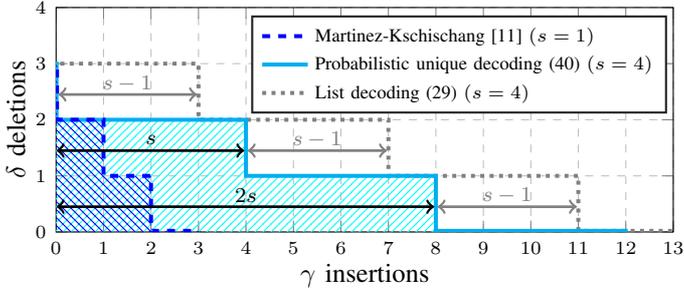

Using similar arguments as in~\cite[Lemma~3]{wachter2014list} it can be shown that the dimension $d_I$ of the $\Fqm$-linear solution space of Problem~\ref{prob:skewIntProblemGenOpLILRS} satisfies
 \begin{equation}
 	d_I\geq \intOrder(\degConstraint+1)-\intOrder k-\insertions.
 \end{equation}
The rank of the root-finding matrix $\RFmat$ can be full if and only if the dimension of the solution space of the interpolation problem $d_I$ is at least $\intOrder$, i.e. if
\begin{align}
 d_I\geq\intOrder\qquad\Longleftrightarrow\qquad
 \insertions+\intOrder\deletions&\leq \intOrder(\nTransmit-k)\label{eq:decRegionLILRSprob}
\end{align}
The probabilistic-unique decoding region in~\eqref{eq:decRegionLILRSprob} is only sightly smaller than the list decoding region in~\eqref{eq:listDecRegion}.
The improved decoding region for~\ac{LILRS} codes is illustrated in Figure~\ref{fig:decodingRegion}.

Combining~\eqref{eq:decDegreeConstraintLILRS} and~\eqref{eq:decRegionLILRSprob}  we get the degree constraint $\degConstraint=\lceil\frac{\nReceive+\intOrder k}{\intOrder+1}\rceil$ for the probabilistic-unique decoder (see~\cite{bartz2017algebraic}). 
Under the assumption that the coefficients $q_{i,j}^{(r)}$ are uniformly distributed over $\Fqm$ (see~\cite[Lemma~9]{wachter2014list}) we have that
\begin{equation}\label{eq:failProbLILRS}
	P_f\leq 4q^{-m(d_I-\intOrder+1)}\leq 4q^{-m\left(\intOrder\left(\left\lceil\frac{\nReceive+\intOrder k}{\intOrder+1}\right\rceil-k\right)-\insertions+1\right)}.
\end{equation}
Our simulations results in the next section indicate that this is also a good estimate of the failure probabiltiy for subspace tuples that are chosen uniformly at random from the set of subspaces from which a given transmitted codeword is ($\insertions,\deletions$)-reachable.
As in \cite{wachter2014list,bartz2018efficient} for interleaved Gabidulin and lifted interleaved Gabidulin codes, respectively, it is an open problem to derive a formal failure probability bound for this non-uniform input. This is subject to future work.
%

%----------------------------------------------------------------------------
% Simulation Results
%----------------------------------------------------------------------------
\section{Comparison to Previous Work and Simulation Results}

The relative overhead due to the lifting is reduced with increasing interleaving order.
Table~\ref{tab:comparisonLLRS_LILRS} shows the improvement of the code rate for increasing interleaving orders.
\begin{table}[ht!]
 \renewcommand{\arraystretch}{1.5}
 \centering
 \caption{Comparison of the dimension $N_i$ of the elementary ambient spaces and the code rate $R$ between LLRS and LILRS codes.}
 \label{tab:comparisonLLRS_LILRS}
 \begin{tabular}{|M{2cm}|M{2.5cm}|M{2.5cm}|}
  \hline
   & LLRS~\cite{martinez2019reliable} & LILRS
   \\ \hline\hline
   Dimension $N_i$ (``packet size'') & $\nTransmitShot{i}+m$ & $\nTransmitShot{i}+\intOrder m$
   \\[8pt] \hline
   Code Rate R & $\frac{mk}{\sum_{i=1}^{\shots}\nTransmitShot{i}(\nTransmitShot{i}+m)}$ & $\frac{\intOrder mk}{\sum_{i=1}^{\shots}\nTransmitShot{i}(\nTransmitShot{i}+\intOrder m)}$
   \\[8pt] \hline
 \end{tabular}
\end{table}

In order to verify the heuristic upper bound on the decoding failure probability in~\eqref{eq:failProbLILRS} we performed a Monte Carlo simulation ($100$ errors) of a code $\liftedIntLinRS{\vecbeta,\a,\shots=2,\intOrder=3;\nTransmit=6,k=3}$ over $\F_{3^3}$ over a multishot operator channel with overall $\deletions=1$ deletion and $\insertions\in\{4,5,6\}$ insertions.

The channel realization is chosen uniformly at random from all possible realizations of the multishot operator channel with exactly this number of deletions and insertions. We implemented this drawing procedure by adapting the efficient dynamic-programming routine in \cite[Remark~7]{puchinger2020generic} for drawing an error of given sum-rank weight uniformly at random.

The results in Figure~\ref{fig:simLILRS} show, that the heuristic upper bound in~\eqref{eq:failProbLILRS} gives a good estimate of the decoding failure probability $P_f$.
\begin{figure}[ht!]
\centering
\definecolor{mycolor1}{rgb}{1.00000,0.00000,1.00000}%
\definecolor{mycolor2}{rgb}{0.00000,1.00000,1.00000}%
\begin{tikzpicture}

\begin{axis}[%
xmin=6,
xmax=10,
xtick={6,7,8,9,10},
xlabel={$\insertions+\intOrder\deletions$},
compat=newest,
xmajorgrids,
ymode=log,
ymin=1e-05,
ymax=1,
yminorticks=true,
label style={anchor=near ticklabel, font=\footnotesize},
label style={inner sep=0}, 
ylabel={Decoding failure probability $P_{f}$},
ymajorgrids,
yminorgrids,
tick label style={font=\scriptsize},
legend style={at={(0.01,0.989)},anchor=north west,legend cell align=left,align=left,draw=white!15!black, font=\footnotesize},
mystyle]

\addplot [color=red,solid,
mark=x,mark options={solid}
]
  table[row sep=crcr]{%
7	2.03e-4\\ 
8	5.48e-3\\ 
9	1.48e-1\\
};
\addlegendentry{Upper bound on $P_{f}$ \eqref{eq:failProbLILRS}};

\addplot [color=blue,
dashed,
mark=o,
mark options={solid}
]
  table[row sep=crcr]{%
7	5.726977e-05\\
8	1.290956e-03\\ 
9	3.883495e-02\\ 
};
\addlegendentry{Simulation $\deletions=1$};

\end{axis}
\end{tikzpicture}%
\caption{Result of a Monte Carlo simulation of the code $\liftedIntLinRS{\vecbeta,\shots=2,\intOrder=3;\nTransmit=6,k=3}$ over $\F_{3^3}$ transmitted over a multishot operator channel with overall $\deletions=1$ deletions and $\insertions=4,5,6$ insertions.}
\label{fig:simLILRS}
\end{figure}
For the same parameters a (non-interleaved) lifted linearized Reed--Solomon code~\cite{martinez2019reliable} (i.e. $\intOrder=1$) can only correct $\insertions$ insertions and $\deletions$ deletions up to $\insertions+\deletions<4$.

\section{Further Applications}

The decoding scheme in Section~\ref{sec:decodingLILRS} can be used to decode \ac{ILRS} codes (without lifting) in the sum-rank metric and \ac{ISRS} codes in the skew metric. 
For the definition of the sum-rank and the skew metric the reader is referred to e.g.~\cite{martinez2018skew}).  

Let the parameters be as in Definition~\ref{def:LILRS} and define $n_i=\nTransmitShot{i}=\nReceive^{(i)}$ for all $i=1,\dots,\shots$. 
Then an $\intOrder$-interleaved linearized Reed--Solomon code $\intLinRS{\vecbeta,\a,\shots,\intOrder;n,k}$ of length $n$ and dimension $k$ is defined as
\begin{equation*}\label{eq:def_ILRS}
  \left\{\!\!\left(\!\!\!
  \begin{array}{c|c|c}
   \opev{f^{(1)}}{\vecbeta^{(1)}}{a_1}  & \!\dots\!\! & \opev{f^{(1)}}{\vecbeta^{(\shots)}}{a_\shots}
   \\[-4pt]
   \vdots  & \!\!\ddots\!\! & \vdots
   \\
   \opev{f^{(\intOrder)}}{\vecbeta^{(1)}}{a_1} & \!\dots\!\! & \opev{f^{(\intOrder)}}{\vecbeta^{(\shots)}}{a_\shots}
  \end{array}
  \!\!\!\!\right)
  \!\!:\!\!\!
  \begin{array}{c}f^{(j)}\in\SkewPolyringZeroDer_{<k}, \\\forall j\in\intervallincl{1}{\intOrder}\end{array}\!\!\!\!\right\}\!\!.
\end{equation*}
Let $\mat{R}=\mat{C}+\mat{E}\in\Fqm^{\intOrder\times n}$ be the received matrix with $\mat{C}\in\intLinRS{\vecbeta,\a,\shots,\intOrder;n,k}$ and the define the interpolation point set 
\begin{equation*}
  \set{P}=\{(\beta_1^{(1)},r_{1}^{(1,1)},\dots,r_{1}^{(\intOrder,1)}), \dots, (\beta_{n_\shots}^{(\shots)},r_{n_\shots}^{(1,\shots)},\dots,r_{n_\shots}^{(\intOrder,\shots)})\}.
\end{equation*}
Then the interpolation-based decoding scheme from Section~\ref{sec:decodingLILRS} can decode up to $t<\frac{\intOrder}{\intOrder+1}(n-k+1)$ errors in the sum-rank metric, which is defined as the sum of the error-rank of each shot.

By using the isometry between the sum-rank metric and the skew metric~\cite{martinez2018skew,martinez2019reliable} we can define an $\intOrder$-interleaved skew Reed--Solomon code $\intSkewRS{\vec{b},\shots,\intOrder;n,k}$ as
\begin{equation}
  \left\{\mat{C}\cdot\diag(\vecbeta^{-1}):\mat{C}\in\intLinRS{\vecbeta,\a,\shots,\intOrder;n,k}\right\}
\end{equation}
where $\diag(\vecbeta^{-1})$ is the diagonal matrix of the vector $\vecbeta^{-1}\defeq\left((\beta_1^{(1)})^{-1},(\beta_2^{(1)})^{-1},\allowbreak\dots,\allowbreak(\beta_{n_\shots}^{(\shots)})^{-1}\right)$ and $\vec{b}\defeq\left(\op{a_1}{\beta_1^{(1)}},\op{a_1}{\beta_2^{(1)}},\allowbreak\dots,\allowbreak\op{a_\shots}{\beta_{n_\shots}^{(\shots)}}\right)\cdot\diag(\vecbeta^{-1})$.
Using the transformation from~\cite[Theorem~9]{martinez2019reliable} the interpolation-based decoding principle from Section~\ref{sec:decodingLILRS} can decode errors of skew weight up to $t<\frac{\intOrder}{\intOrder+1}(n-k+1)$.

%----------------------------------------------------------------------------
% References
%----------------------------------------------------------------------------

\clearpage
\enlargethispage{-2.5cm} 
\bibliographystyle{IEEEtran}
% \bibliography{references}

\end{document}